\documentclass[journal]{IEEEtran}
\usepackage{lineno} 
\modulolinenumbers[5]
\usepackage{booktabs} 
\usepackage{amsmath,amsthm}
\usepackage{amsfonts,amssymb}
\usepackage{algorithm,algorithmic}
\usepackage{graphicx}  
\usepackage{subfigure}
\usepackage{multirow}
\usepackage{enumitem}
\usepackage{xcolor}
\theoremstyle{definition}

\begin{document}
\title{Network Security in the Industrial Control System: A Survey}

\author{Yang Li, 
Shihao Wu,
Quan Pan
\thanks{All of the authors are with the School of Automation, Northwestern Polytechnical University.}
\thanks{Please note that all of the works we summarized are before 2020.}}


\maketitle
\begin{abstract}

Along with the development of intelligent manufacturing, especially with the high connectivity of the industrial control system (ICS), the network security of ICS becomes more important. And in recent years, there has been much research on the security of the ICS network. However, in practical usage, there are many types of protocols, which means a high vulnerability in protocols. Therefore, in this paper, we give a complete review of the protocols that are usually used in ICS. Then, we give a comprehensive review on network security in terms of Defence in Depth (DiD), including data encryption, access control policy, intrusion detection system, software-defined network, etc. Through these works, we try to provide a new perspective on the exciting new developments in this field.

\end{abstract}


\section{Introduction}
\label{sec:introduction}
The new generation of information technology, especially the new generation of intelligent manufacturing, is developing rapidly and accelerating its integration with the internet, which brings new opportunities for the transformation and upgrading of the global manufacturing industry.
However, the high connectivity of industrial control systems (ICS) makes their security an important issue. In particular, the diversity of industrial control network protocols increases the vulnerability.
More than 70\% of the vulnerabilities disclosed in ICS in the first half of 2020 were remotely exploited by cyber attack carriers, according to an industrial cybersecurity firm~\footnote{https://www.securityweek.com/over-70-ics-vulnerabilities-disclosed-first-half-2020-remotely-exploitable}.

In the view of cybersecurity, network security is confidentiality and non-repudiation of the communication in ICS, which includes protocol security, network structure security, etc. 
To ensure network security for ICS network designs, different organizations come up with different standards.
For example, the National Institute of Standards and Technology (NIST) proposed the guidelines for ICS Security~\cite{stouffer2015guide} since 2011, 
the International Electrotechnical Commission (IEC) proposed the ISA/IEC 62443-4-1~\cite{international2018iec} in 2018 to ensure lifecycle security in ICS, etc.
Different from traditional networks, industrial devices can be divided into different sectors or zones according to their functions and positions in ICS.
Therefore, defense-in-depth (DiD) is an important way to ensure network security for the entire industrial network system. 
A DiD, usually includes data encryption, access control policies, intrusion detection system, etc.
Data encryption is to ensure the confidentiality of data transmission, usage, and storage.
The access control policy is a direct way to protect the ICS from hostile detection. 
And the intrusion detection system is a mostly used way to monitor malicious activity or policy violations in the ICS.
 Therefore, in this survey, we will elaborate on how these strategies work and how they typically behave in an ICS system. 
 
Plenty of research has been done on ICS network security. However, as far as we know,  there are very few systematic reviews that well shape this area and current progress. Although some works have given the spotlight on the survey of ICS security, e.g.,  
Knowles et al.~\cite{knowles2015survey} surveyed the methodologies and research that all before 2015 in the view of ICS security measuring and risk management. 
Xu et al.~\cite{xu2017review} reviewed the works in view of the protocols that are used in ICS.
Recently, You et al.~\cite{you2018review} provided a brief introduction to ICS security given security control tendency, ICS operation, network layer, etc.
Although these works have explored ICS security from different views, there is still lacking a systematic survey on the current progress of network security in ICS.
Furthermore, different from previous studies, our focus is more on the defense in depth which is more practical.

The structure of this paper is described as follows: In section~\ref{sec:com}, we provide a discussion about the ICS network. In section~\ref{sec:def}, we give a discussion about the defense in depth which expounds on data encryption, access control policy, intrusion detection, software-defined network, etc. Finally comes the future research directions in section~\ref{sec:future}


\section{ICS Networks}
\label{sec:com}
With the development of connectivity and openness, communication security has become a major threat to ICS. 
In view of different scenarios, the protocols, network structure, etc., may be different in an ICS. 
Meanwhile,  one of the features of ICS is the diversity of protocols and there is no unified standard for its design. 
Therefore, it will be a challenging task to guarantee network security in ICS.
In this section, we will describe the protocols, and network structure in ICS.
\subsection{Network Structure} 

An ICS network is organized by: Programmable Logic Controller (PLC) which is an industrial digital computer used in controlling manufacturing processes, Remote Terminal Unit (RTU) which is a microprocessor-controlled electronic device that interfaces objects in the physical world to the SCADA system, Intelligent Electronic Device (IED) which is an integrated microprocessor-based controller that usually used in power system, and Human Machine Interface (HMI) which is a control panel that operates PLCs, RTUs.

\begin{figure}[htp]
   \centering
      \includegraphics[width=0.96\linewidth]{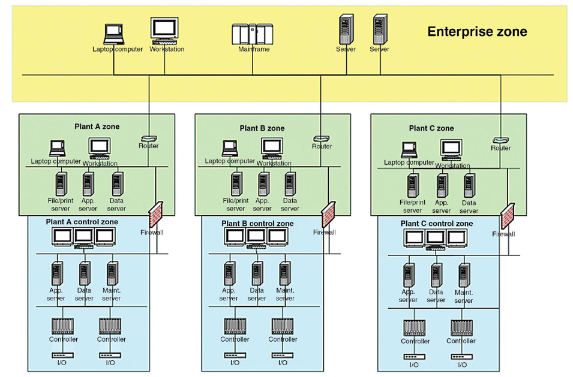}   
    \caption{An example of the zone in an ICS network (ISA-62443 zone), cited from~\cite{knapp2014industrial}.}
    \label{fig:example}
\end{figure}

\begin{figure}[th]
   \centering
      \includegraphics[width=0.96\linewidth]{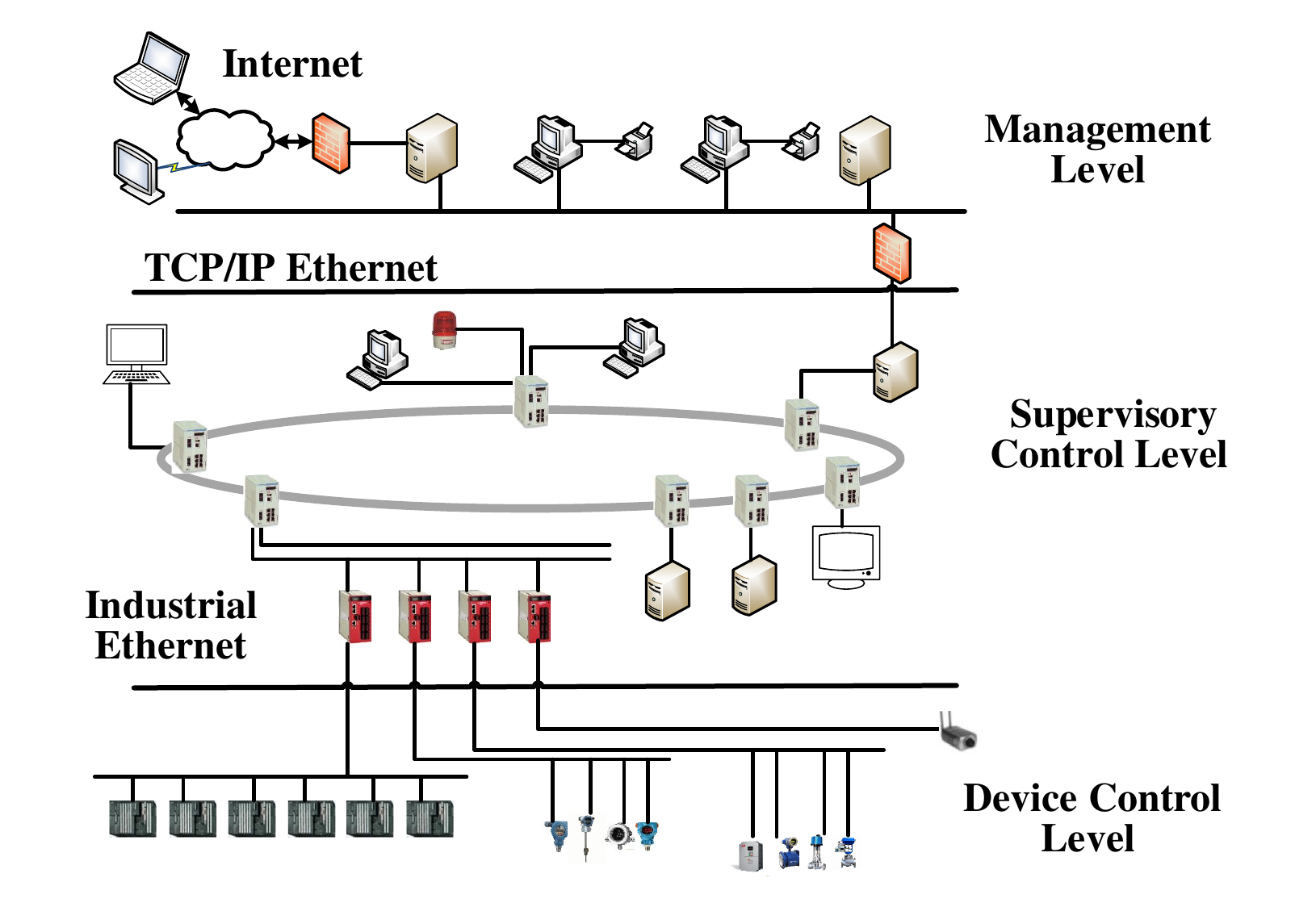}   
    \caption{Another example of the ICS topology.}
    \label{fig:defenseindepth}
\end{figure}

Based on the different requirements of ICS application scenarios,  different network topologies are needed. 
For example, the devices are usually divided into \textit{groups}, or \textit{zone} based on their location, usage, function, or network security~\cite{knapp2014industrial}, which is also recommended by NIST~\cite{stouffer2015guide}, details will be introduced in Section~\ref{subsec:zone}.
As recommended in the standard ISA-62443~\footnote{https://www.isa.org/training-and-certification/isa-certification/isa99iec-62443/isa99iec-62443-cybersecurity-certificate-programs}, the topology can be designed as the enterprise zone and plant zones in terms of the network security and functions which is illustrated in Figure~\ref{fig:example}.
To ensure security, there is no connection among zones.
Another commonly used topology, shown in Figure~\ref{fig:defenseindepth}, is a three-level ICS network: management level, supervision control level, and device control level. The management level can be seen as a traditional IT structure whose security is guaranteed by traditional defense methods. And in the supervision control level and device control level, different kinds of protocols are applied, e.g., ModBus~\footnote{https://modbus.org/}, ProfiNet~\footnote{https://www.profibus.com/technology/profinet/}, DNP3~\footnote{https://www.dnp.org/About/Overview-of-DNP3-Protocol}, etc. However, seldom security modules are applied in those two levels in practice due to the incompatibility of manufacturing devices and protocol diversity. 

Previously, cable distance, traffic distribution, traffic balance, network delays, etc., are the main considerations in the ICS network designing~\cite{zhou2010study}, especially in the real-time system. 
However, with the interconnectivity of ICS, network security has become a serious issue. Also, the researchers start to focus on safe neural network designing in ICS. But it is hard to reconcile the traditional requirement and the security. The general problems that an ICS network needs to concern with include real-time data transfer, geographical position limitation, strong determinism, etc., which are difficult to harmonize, and the security requirement makes it even harder~\cite{genge2017cyber}.

\subsection{Protocols}

The protocol is the basic element for communication in ICS, and plenty of works focused on the security of a specific protocol. And the truth is that massive protocols exist in ICS and most of them are non-public.
More sadly, most of the protocols do not consider the security mechanism when they are designed and applied. 
In this subsection, the commonly used protocols in ICS will be analyzed in authorization, encryption, availability, integrity, and confidentiality those aspects respectively\cite{xu2017review}.
The details are listed in Table~\ref{tab:protocols}.

\begin{table*}[ht]
\centering
\begin{tabular}{l|c|c|c|c|c|c}
\hline
                & \textbf{Authentication}   & \textbf{Authorization}  & \textbf{Encryption} & \textbf{Availability} & \textbf{Integrity} & \textbf{Confidentiality}   \\\hline
Profinet~\cite{muller2018profinet,niemann2019security}  &                            &         $\surd$                 &         $\surd$                &       $\surd$                       &             $\surd$            &                             \\\hline
DNP3~\cite{darwish2015smart,siddavatam2015security,amoah2016securing}     &                             &                        &                        &     $\surd$              &                    &   $\surd$                  \\\hline
Modbus~\cite{nardone2016formal,phan2012authenticated}   &                            &          $\surd$         &      $\surd$             &                         &                    &    $\surd$                         \\\hline
IEC 60870-5104~\cite{yang2013intrusion,maynard2014towards}  &                 &         $\surd$          &      $\surd$             &                        &                    &    $\surd$                         \\\hline
IEC 61850~\cite{youssef2016iec,kabir2016test}        &                   &                       &                      &       $\surd$             &        $\surd$          &    $\surd$                         \\\hline
IEC 61400-25~\cite{nguyen2012smart,liu2008security}   &       $\surd$         &          $\surd$         &        $\surd$          &        $\surd$           &                    &    $\surd$         \\\hline
IEEE C37.118~\cite{khan2016analysis}   &         $\surd$         &          $\surd$         &        $\surd$          &                      &                      &       \\\hline
\end{tabular}
\caption{The detail of the weakness in common used ICS protocols (Partly cited from~\cite{xu2017review}).}
\label{tab:protocols}
\end{table*}

In the table, \textbf{Authentication} indicates the mechanisms that judge an identity, if there is no authorization in a protocol, the privileges can be easily gained with the forge protocol packets.
\textbf{Authorization} is a security mechanism to determine access levels based on the user's identity. If there is a lack of authorization, malicious users can send any information or resource to others without permission.
\textbf{Encryption} denotes whether there is cryptography in the protocol, if there is no encryption, the communication data can be captured easily by malicious attackers. \textbf{Availability} is the status of the ICS device or service, if it is a lack of availability, ICS may lose control which may cause an industrial accident. \textbf{Integrity} indicates whether the data is transmitted completely, if there is a lack of integrity, communication data can be rendered useless by package missing or corruption. 
\textbf{Confidentiality} refers to ICS's protection against unauthorized access and misuse. Without confidentiality, unauthorized users can exploit vulnerabilities to achieve illegal purposes.

Real-time Ethernet (RTE) is a set of protocols that support real-time operation, and all of those protocols are designed based on the IEC 61784-1. All protocols mentioned below belong to RTE.

\subsubsection{Profinet}
Profinet is usually used in data communication between controllers (e.g., PLCs, DCSs, or PACs, etc.) and devices (e.g., I/O blocks, RFID readers, proxies, etc.). 
There are four layers in Profinet which are Ethernet (physical and data link layers), IP (network layer), TCP\&UDP (transport layer), and other protocols (application layer). 
In some specific Profinet versions, to make Profinet faster, it may skip the IP, and TCP\&UDP layers, e.g., Real-Time Profinet. 
Profinet has two versions which are Profinet CBA and Profinet IO,  Profinet CBA is suitable for component-based communication via TCP/IP, while Profinet IO is used in systems requiring real-time communication. 
As there is no authentication mechanism in the original version.
Therefore, it suffers from the network security problems, such as man-in-the-middle, including packet flooding, packet sniffing, etc. 
To tackle this problem, Oh et al.~\cite{oh2014advanced} added an Integrity Check Value (ICV) module at the base of Profinet/DCP (with DCP as configuring protocols) to the authentication data field.
Guilherme et al.~\cite{sestito2018method} proposed an anomaly detection method that can identify four different events that happened in Profinet.
In summarizing, there is a high authorization, encryption, integrity, availability w.r.t., network security~\cite{muller2018profinet,niemann2019security}.

\subsubsection{DNP3}
DNP3 is designed based on the IEC TC-57 standard, and it is a three layers protocol. Even with the large scale of the application, DNP3 is still weak in security guaranteed. Firstly, there is no authentication protection in DNP3, and it is easy for an attacker to disrupt the control process by creating a normal conversation. Secondly, it is also lacking authorization protection, and any user can run any functions over it. Finally, there is no encryption protection, and the messages it transmits are in plain text. To ensure the DNP3's security,  Jeong-Han et al.~\cite{yun2013burst} detected the intrusion by producing a burst-based whitelist model. Bai et al.~\cite{bai2014network} proposed an automation protection framework to detect the attacks that aim at the DNP3. Hao et al.~\cite{li2015designing} designed a set of snort rules on the DNP3 networks for intrusion detection. Sahebrao et al.~\cite{shinde2018cybersecurity} enhanced the DNP3's security by modifying the internal protocol structure and encrypting its packages with Blowfish~\cite{schneier1993description}, and Ye et al.\cite{lu2018research} proposed an improved version of DNP3-BAE by using the hashing chain. 

\subsubsection{Modbus}
Modbus is one of the most popular and oldest protocols in the ICS communication module, it was designed to connect with PLCs in 1979, and now it is broadly applied in industries with two kinds of implementation. One is serial Modbus, which applies the high-level data link control (HDLC) standard\footnote{http://en.Wikipedia.org/wiki/High-Level\_Data\_Link\_Control}. And the other one is the Modbus-TCP, which adopts the TCP/IP protocol stack~\cite{span2015}. Same to DNP3, lots of works are proposed to enhance this protocol. Emil et al.~\cite{pricop2017method} proposed an authentication method for the device when connecting with Modbus TCP. Fei et al.~\cite{fei2018security} designed a new format of Modbus named SecModbus, it did not increase the communication procedure and ensured confidentiality, integrity, and authorization at the same time. 
Apart from the improvement over the Modbus protocol, Yusheng et al.~\cite{yusheng2017intrusion} designed the stereo depth intrusion detection system (SD-IDS) to inspect the Modbus TCP traffic.

\subsubsection{IEC 60870-5104}

IEC 60870-5104 (also known as IEC 870-5-104) is an international standard, released in 2000 by the IEC (International Electrotechnical Commission). 
Several protocols are designed based on this standard, e.g., ip4Cloud/SEC3PB which is to capture Profibus data by eavesdropping and transmitting it to  SCADA services, and ip4Cloud/SEC3IO which is to switch and monitors digital IO status to transmit them to SCADA services. 
The protocol under this standard can be used for remote control tasks between the control center and the substation.
 Generally, there is no authentication and encryption mechanism in IEC 60870-5104, and all the messages are transmitted with plain text. 
 Due to such problems, protocols under this standard are vulnerable:
 Maynard et al.~\cite{maynard2014towards} analyzed the attack behaviors to the IEC 60870-5104 protocols and deployed the man-in-the-middle attack over the SCADA system successfully and suggested that rule-based methods can be applied in the security insurance. 
 Recently, Qassim et al.~\cite{qassim2018simulating} showed that a successful control command injection attack can be implemented by exploiting the previously identified vulnerabilities in their designed SCADA testbed.
To ensure security,  with the help of the open-source Snort tool\footnote{https://www.snort.org/}, Yang et al.~\cite{yang2013intrusion} used a rule-based method to detect the intrusion that deployed on this protocol.
To protect the substation automation network based on IEC 60780-5104 protocol, Hodo et al.~\cite{hodo2017anomaly} proposed a machine learning-based framework to do intrusion detection.

\subsubsection{IEC 61850}

IEC 61850 is a standard developed by IEC Technical Committee no. 57 Working Group 10 and IEEE for Ethernet (IEEE 802.3) based communications in substations. It is an international standard defining communication protocols for intelligent electronic devices at electrical substations. 
As with many other communication standards, IEC 61850 was developed without extensive consideration of critical security measures.
It always suffers from plenty of attacks including eavesdropping, data spoofing, DOS, password cracking, etc~\cite{weerathunga2012security}.
To tackle the security problems in IEC 61850, IEC 62351 (Part 3, 4, and 6) is developed by the same committee in 2007. For details please refer to Youssef et al.'s work~\cite{youssef2016iec}. 

\subsubsection{IEC 61400-25}
 IEC61400-25 is a standard developed by the IEC TC88 Technical Committee. 
 It is specifically aimed at the monitoring system communication of wind power plants. 
This standard aims to realize free communication between equipment of different suppliers in wind power plants~\cite{ahmed2014hierarchical}. 
The security of network communication between wind power plants is required in IEC 61400-25-3. But the method of implementation is not specified, and it is left to the protocol itself~\cite{international2006wind}.
With the development of the Internet, wind power plant communication based on Web services has become a common way.
However, there are significant potential security risks, and while access can be restricted by user name/password, they do not provide additional protection for confidentiality and integrity.
To ensure the security of Web-based service,  Liu et al.~\cite{liu2008security} extended the simple object access protocol message, and designed the security agent with a security message processing algorithm to achieve confidentiality integrity and authentication across the communication.

\subsubsection{IEEE C37.118}

IEEE C37.118 defines the data transmission format and synchronization requirements between the communication of substations. 
This protocol defines four different types of messages, including data, header, configuration, and command.  
Like most other protocols, it lacks predefined security mechanisms, which makes it vulnerable to network attacks.
Attacks like reconnaissance, man-in-the-middle, DOS, etc., severely impact the synchrophasor application, sometimes, it may even cause physical damage to the equipment~\cite{khan2016ieee}.
To ensure the security of this protocol, Stewart et al.~\cite{stewart2011synchrophasor} validated the security by using the firewall and Virtual Private Network between the substation and control center. 
Basumallik et al.~\cite{basumallik2018cyber} recommended encryption algorithms (e.g., AES), IPSEC, and DTLS  to ensure the confidentiality and authentication of the data without affecting the instantaneity.




\section{Defense in Depth  }
\label{sec:def}
It is a widely-known concept in IT security with defense in depth (DiD). To make a redundant security defense system in ICS, DiD also is necessary along with the development of ICS's connectivity. 
It is better to apply DiD in all ranges of ICS that include proactive protection, such as data encryption and access control, etc., as well as reactive protections, such as intrusion detection, intrusion protection, and network range, etc. In this section, all of these topics will be introduced here.

 \subsection{Data Encryption}
\begin{figure}[th]
    \begin{center}
      \includegraphics[width=0.85\linewidth]{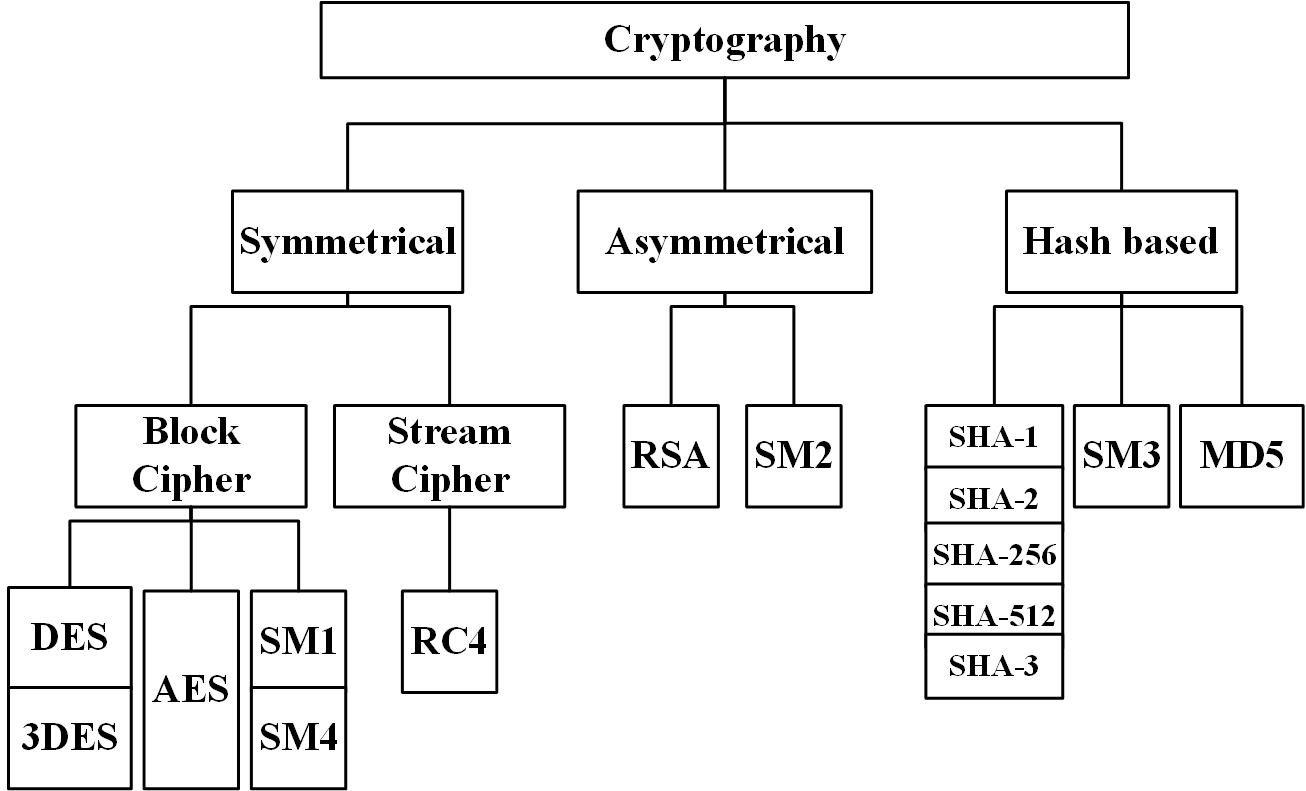}
    \end{center}
    \caption{The structure of the cryptography.}
    \label{fig:structure_cryptography}
\end{figure}

Encryption which is to transform the plaintext data to cipher data using the cryptographic function is an important way to ensure the confidentiality and integrity of the data.
As we know, encryption is the fundamental way to guarantee the security of the data. Encryption has been a necessary method to ensure the security of ICS.
 The cryptographic methods can be divided into those three categories, the first one is symmetrical cryptography which uses the same key for the transformation, the second one is asymmetrical cryptography (also named public-key encryption) which using the public and private keys separately when doing the encryption and decryption, the public key can be open, while the private key needs to be kept secret. And the third one is hash-based cryptography which mainly is applied in integrity verification. The technique structure of the cryptographic methods shows in Figure~\ref{fig:structure_cryptography}

In cryptography, there are two different kinds of cipher methods, one is a block cipher which encrypts a block of text at a time, rather than encrypting a bit that doing in the stream cipher. Block cipher is safer than the stream cipher, but it cost more time during the encryption. Methods like DES~\cite{coppersmith1994data}, 3DES~\cite{coppersmith1996proposed}, AES~\cite{daemen2001reijndael}, SM1 and SM4, etc., are all the block cipher, which usually are deployed in the communication system in ICS. 
However, a stream cipher is fast but it is easy to be cracked. The most popular stream cipher is RC4 which was used in Transport Layer Security (TLS) before 2015~\cite{popov2015prohibiting}. As to asymmetrical cryptography, The RSA and SM2 belong to asymmetrical cryptography, and SHA-series, SM3, and MD5 are hash-based cryptography. Based on the applications, different cryptographies can be deployed. 


In ICS, before deploying the cryptographic algorithms, such things should be taken into consideration: high safety, low cost, and high performance~\cite{shanshanxu2016}. However, due to the special need in the ICS, there are different voices in encryption applications in ICS. 
Fauri et al.~\cite{fauri2017encryption} made a critical discussion about the encryption application in ICS, and concluded that the encryption process will not help increase the overall safety of the ICS, sometimes the encryption process even has negative consequences for the security. 
Therefore, different cryptographic algorithms or standards are applied along with different applications. 
For example,  In the power systems infrastructure, IEC 62351 recommends end-to-end protocol TLS and point-to-point protocol IPsec, however, it is end-to-end protocol WSSeurity in the industrial automation system.

\subsection{Access Control Policy}
It is a proactive protection method for ICS security, which includes a firewall, network address translation (NAT), etc. It is a direct way to make the access control policy to protect the ICS from hostile detection. 
 
\subsubsection{Firewall}
A common method is adopting the white list in the firewall to prevent unknown access~\cite{yoo2013whitelist}. Byoung-Koo et al.~\cite{kim2016abnormal} designed the Industrial Cyber Attack Prevention-Gate system for the ICS, and aimed to prevent unauthorized access fundamentally by applying the firewall in all of the levels mentioned before, the drawback was decreasing the latency of the ICS in return. To reduce the latency of the firewall matching in the white list, the hash-based rule lookup was proposed by Pabilona et al.~\cite{lee2017enhancing}

There are two ways to white list building, one is the static list, and the other one is the generated-dynamic list. The static list is time costing, inflexible, and limited in expression, but it is precise in prevention. And it is the opposing situation in the generated-dynamic list. There are lots of works in exploring static list building. The distributed firewall was proposed by Peng et al.~\cite{peng2012analysis}, and each device was configured with different policies by using the static white list. Woo-Suk et al.~\cite{jung2017structured} applied the PrefixSpan algorithm to generate the structured static white list, which improves the flexibility of the static list. Besides the static white list, plenty of works are proposed for dynamic list generation. Barbosa et al.~\cite{barbosa2013flow} learned the white list via the flow by using the dynamic port allocation. Choi et al.~\cite{choi2015traffic} generated the white list automatically based on the flow's locality. There also are some works learning the white list with dynamic packet inspection. Jeyasingam et al.~\cite{nivethan2016linux} extended the Linux-based firewall for the DNP3 protocol in the power grid, and the u32 feature-match mechanism lets the firewall extract any parts of the package, which made the white list more dynamic.

\subsubsection{Security Zone}
\label{subsec:zone}
Based on the standard proposed by NIST~\cite{stouffer2015guide}, the ICS network usually partitions into several different zones (such as trust zone, demilitarized zone, etc.), especially for some sensitive companies which have high requirements for security.
However, lots of problems need to deal with. Bela et al.~\cite{genge2017cyber} treated the network designing in ICS as the integer linear programming (ILP) problem, and the security are added as the constraints to fulfill security requirement, finally, the results are assessed by the cyberattack impact assessment (CAIA). Jun et al.~\cite{yang2018anomaly} proposed an automated zone partition method based on the physical system causal model to do anomaly detection. This method adopts the zones crucial states as the input, which means more data should be packaged back into the system to make the zone partition decision.

\subsection{Intrusion Detection System}

\subsubsection{Vulnerability Detection}
\begin{figure}[th]
    \begin{center}
    \subfigure[The vulnerabilities number discovered in ICS since 1997]{
      \includegraphics[width=0.45\linewidth]{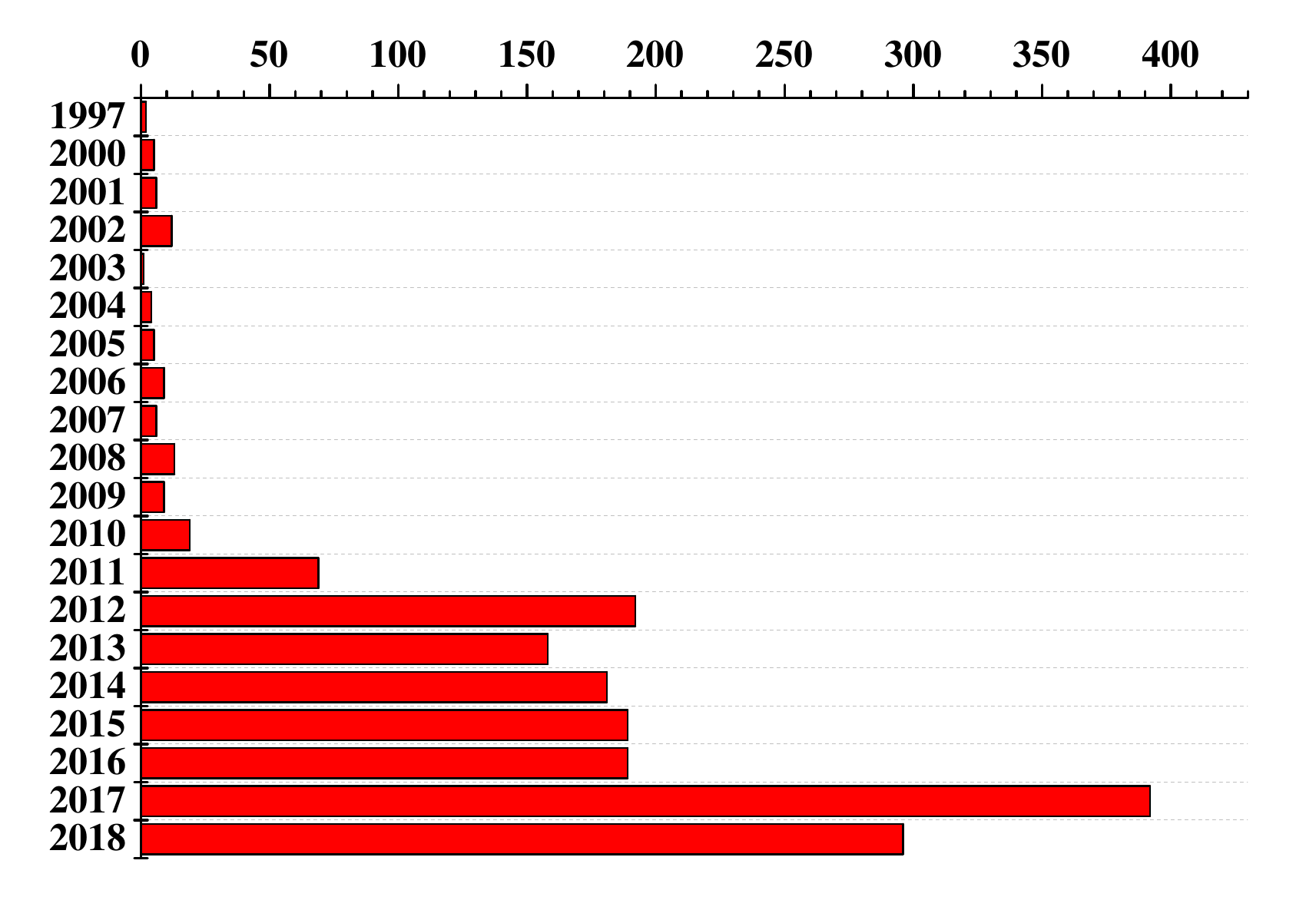}}
      \subfigure[The vulnerabilities distribution in year of 2018 by the risk level]{
      \includegraphics[width=0.45\linewidth]{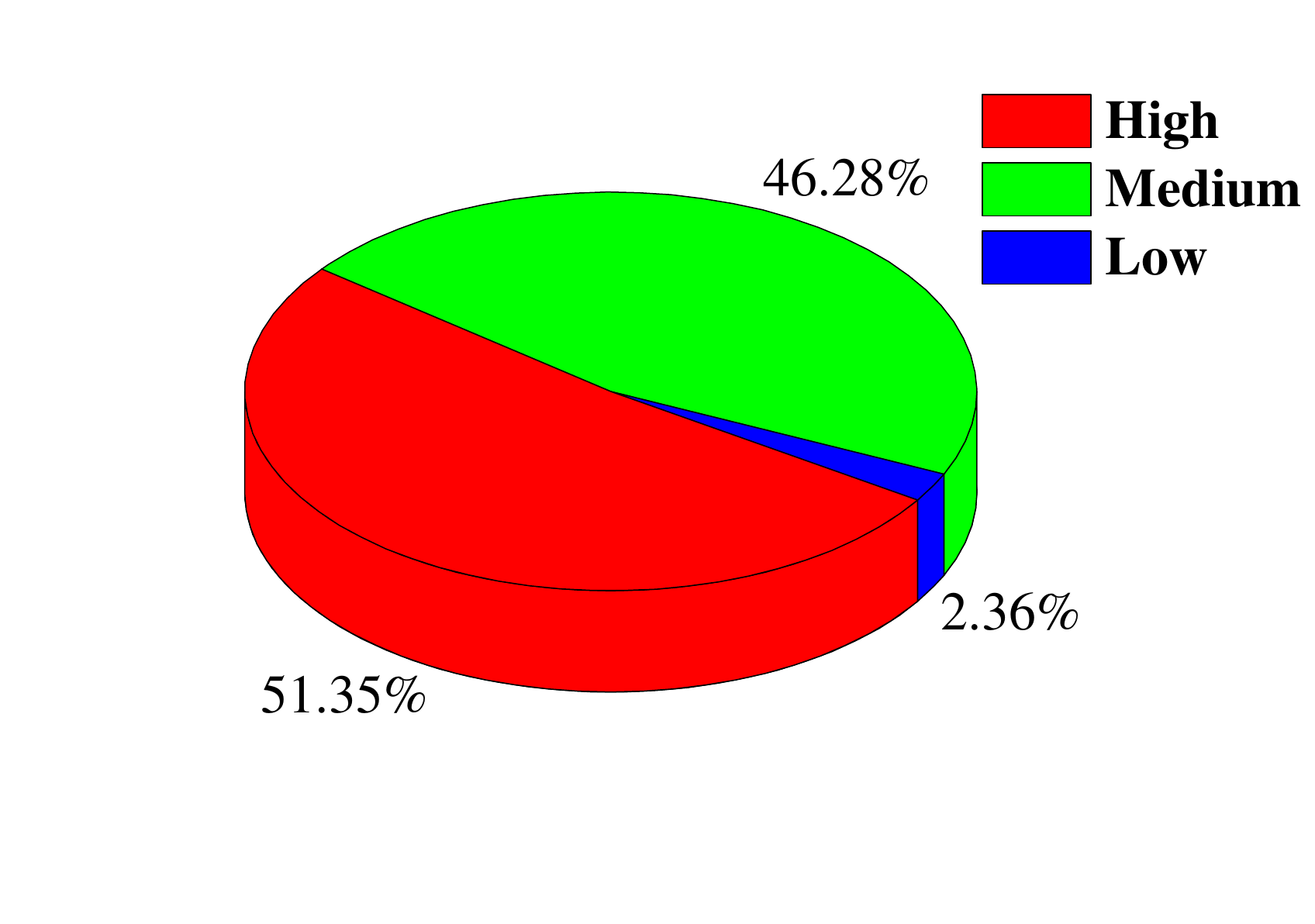}}
    \end{center}
    \caption{The vulnerabilities distribution in ICS.}
    \label{fig:vum_num}
\end{figure}

Vulnerabilities are a common problem in traditional systems, as well as in ICS.  The number of vulnerabilities in ICS has increased dramatically since they were first published in 1997~\cite{andreeva2016industrial}, especially in the recent seven years which can be seen in Figure~\ref{fig:vum_num}(a). Before 2015, the number is cited from~\cite{andreeva2016industrial}, and after 2015, the numbers are counted from the website of CNVD~\footnote{http://ics.cnvd.org.cn/})\footnote{http://ics.cnvd.org.cn/}, the number of vulnerabilities in 2018 is counted before 10th September.
From Figure~\ref{fig:vum_num}(b), we can see that, among the vulnerabilities published in 2018, most of them are at high-level risk for the ICS.
 Only 2.36\% of them are in low-level risk, which has the same trend that reported by Oxana et al.~\cite{andreeva2016industrial}. 

 Vulnerabilities can be exploited in any place of ICS, not only in its software or firmware but also in tools that are associated with monitoring and auditing processes. And most of the vulnerabilities are unknown. 
The components of vulnerability, it can be divided into four types: system platform vulnerability, communication module vulnerability, application software vulnerability, and hardware vulnerability.
 The system platform is to provide the base service for the ICS, which always refers to SCADA in ICS. The communication module is the communication system in ICS, which includes the communication protocols, bus, and I/O system, etc. Application software denotes the software that associates with the SCADA function which includes the management, production, storage, and operation. Hardware is the basic component that supports the ICS, which is the device that includes the circuits, RTU, etc. 
According to the function of vulnerability, vulnerability can also be divided into buffer overflow, authentication bypass, cross-site vulnerability, and sensitive information sniffing four categories.
A buffer overflow is an overflow of the legal boundary of a buffer caused by a programming error.
This type of vulnerability has been discovered in all of the four components mentioned before. 
Authentication bypass is the vulnerability that allows an attacker to gain access to information without authentication and is always present in the system platform, communication module, and application software.
The cross-site vulnerability that allows an attacker to inject insecure scripts into a server is always present on a system platform, e.g., SCADA.
Sensitive information sniffing is the vulnerability that allows an attacker to obtain or remove sensitive information from ICS. This vulnerability is always present in the communication module.
These four categories are not independent, and sometimes these categories of vulnerabilities form a so-called \textit{chain} vulnerability, giving an attacker more opportunities to attack ICS~\cite{xu2017identification}.

Vulnerability modeling is an important work in vulnerability detection. One simple detection way is to use a vulnerability database.
A number of vulnerability databases are available online, such as Common Vulnerabilities and Exposures (CVE)\footnote{http://cve.mitre.org/}, Open Source Vulnerability Database (OSVDB)\footnote{http://osvdb.org/}, National Vulnerability Database (NVD)\footnote{http://nvd.nist.gov/}, China National Vulnerability Database (CNVD)\footnote{http://ics.cnvd.org.cn/}, Security Focus’s vulnerability database\footnote{https://cirdb.cerias.purdue.edu/coopvdb/public/}, and Public Cooperative Vulnerability Database (PCVD)\footnote{http://www.securityfocus.com/vulnerabilities}, etc. 
However, detection with a vulnerability database has many limitations, including description shortage in vulnerabilities' presence, exploit-ability, and effect, unreadable to machine etc.~\cite{maggi2008vulnerability}
Faced with these problems, Sufatrio et al.~\cite{yap2004machine} proposed the newly designed vulnerability database Movtraq, which is machine-readable and can be directly applied to automated detection systems.
But this database still has some drawbacks:
First of all, Movtraq relies on Unix systems (RedHat and FreeBSD) to run and is not portable;
Second, the information it focuses on is unitary, which makes it difficult to make a further application.
Therefore, we urgently need a language that can describe vulnerabilities completely.  Open Vulnerability and Assessment Language (OVAL)\footnote{http://oval.mitre.org/index.html} is a machine-readable XML-based language that is defined by MITRE\footnote{https://www.mitre.org/}.
There are three parts in OVAL which are system information representation, machine state expression (e.g., vulnerability, configuration, patch state, etc.), and assessment results~\cite{banghart2010open}. 
The vulnerabilities described by OVAL can be entered directly into the scanner, but they have no information about their exploitability and need to be checked manually. 
Based on OVAL, new vulnerabilities modeling language DEpendability and Security by Enhanced REConfigurability (DESEREC)\footnote{http://www.deserec.eu/}~\cite{perez2007deserec} is designed. 
This language can effectively describe the exploitation of vulnerabilities and has been successfully applied to the automatic vulnerability detection system~\cite{cheminod2009detecting}. 

As we know, there are lots of vulnerabilities scanning tools, such as SARA\footnote{http://www-arc.com/sara/}, SAINT\footnote{http://www.saintcorporation.com/} and Nessus\footnote{http://www.nessus.org} etc. 
In addition to these tools, search engines such as Google\footnote{http://www.google.com} and Shodan\footnote{https://www.shodan.io/} can also be applied in vulnerabilities detection~\cite{simon2016vulnerability}. 
But their limitation is that they provide few clues, especially when faced with exploiting serial vulnerabilities on multiple hosts~\cite{maggi2008vulnerability}.
To deal with such a situation,  Manuel et al.~\cite{cheminod2009detecting} extended the OVAL by introducing two new elements which are \textit{preconditions} and \textit{postconditions}. 
Recently, Cheminod et al.~\cite{cheminod2015analysis} have applied high-level security policies to modeling to make vulnerability detection more effective.
Generally, the vulnerabilities that these systems deal with are mainly in the system platform, application software, and communication module. 

In addition to vulnerability modeling, there are other methods for vulnerability detection, such as virtual technology, Fuzzing test, etc.
Ashlesha et al.~\cite{joshi2005detecting} designed a system IntroVirt based on vulnerability-specific predicates in virtual-machine introspection to detect vulnerabilities. IntroVirt can detect or respond to past and present vulnerabilities.
Xiong et al.~\cite{xiong2015vulnerability} using the Fuzzing test to detect the vulnerabilities in Modbus-TCP.
Kim et al.~\cite{kim2018field} proposed a test case generation technique for a fuzzing test that can be used for vulnerability detection in industrial control system protocols. 
Luo et al.~\cite{luo2019polar} proposed a functional code-aware fuzzy identification framework Polar, which can automatically extract semantic information from the ICS protocol and use this information to accelerate vulnerability detection.

\subsubsection{Malware Detection}

Unlike the passive threat invulnerability, the malware threat is active and can be extremely disruptive to the system, such as Stunex~\cite{karnouskos2011stuxnet}. 
We know that malware has been discovered since the invention of computers.
The openness of ICS makes it more vulnerable to traditional malware infections.
Andrea et al. ~\cite{carcano2008scada} have shown that ICS can be infected even without customized malware.
Malware refers to software or code intended to read, write, or change the normal state of the ICS. It includes viruses, worms, Trojans, malicious code, etc.
Malware usually exploits vulnerabilities in ICS.

When doing malware detection in ICS, ``real-time'' is an important factor that needs to concern, especially when ICS is in a production state. 
Some works divide malware detection into two categories, one is anomaly-based detection and the other one is signature-based detection~\cite{idika2007survey,zhu2010scada}. 
Anomaly-based detection should know the normal and the abnormal behaviors that malware may act on~\cite {yang2006anomaly}. 
This approach requires the detection tool to know all the normal behavior of the software in ICS.
Signature-based detection, however, usually tries to find common features of malware.
This detection method can usually be implemented by machine learning, rule-based systems, checksums, scanning strings, etc.~\cite{bist2012classification,peng2016malware}. 
Machida et al.~\cite{machida2020novel} applied rule-based methods to lure this malware into our sensors by continuously embedding sensor information into the host list in the ICS network, and it successfully detected malware like WannaCry and Conficker.
 In addition to these methods, formal languages can also be used in malware detection.
Saman et al.~\cite{zonouz2014detecting} detects the malware automatically in the PLC through code analytics with formal language. Generally, this method belongs to signature-based detection but with specific-defined malware states informal language. 

Anti-virus simulators were usually applied in the past, and they detected the malware through the simulation of the behavior, such as Hirst's Virus Simulation Suite~\cite{hirst1990virus}, Virlab~\cite{faistenhammer1993virlab}, Nepenthes~\cite{baecher2006nepenthes} etc. But all of those projects are out of date and stopped updating ten years ago. 
Luis et al.~\cite{garcia2017hey} designed the HARVEY system embedded in the PLC to detect the malicious command that malware sends to PLC.

\subsubsection{Anomaly Traffic Detection}
Another problem the ICS always faces is anomaly traffic, which is a kind of intrusion over industrial networks. Because of the complexity of the ICS, it is more difficult to do intrusion detection in the ICS than that on the traditional Internet.

Intrusion Detection is a communication security technique that can detect, identify, and respond to unauthorized operations, such as insert, delete, query, and modify. There are three main principles for intrusion detection: misuse-based, anomaly-based, and hybrid(combine the formers)~\cite{7307098}.

The misuse-based method can be used to detect known attacks by the signatures of these attacks. This technique can detect the known type of attacks effectively, but it needs to manually update the database frequently, and can't detect novel attacks.

The anomaly-based method can identify anomalies from normal behavior by modeling the normal network and system behavior. This method can detect unknown attacks, but it has high false alarm rates because the unseen behaviors may be identified as anomalies.

The hybrid method combines the misuse-based method and the anomaly-based method and inherited their respective advantages. This method can decrease the false positive(FP) rate for unknown attacks, and raise the detection rates of known attacks.

The emergence of machine learning approaches makes Intrusion Detection face a new opportunity, these approaches learn from the available data and mine the unknown characteristics in the data. At present, The following machine learning and data mining methods can be used in Intrusion Detection: Artificial Neural Networks, Association Rules and Fuzzy Association Rules, Bayesian Networks, Clustering, Decision Trees, Ensemble Learning, Evolutionary Computation, Hidden Markov Models, Inductive Learning, Naive Bayes, Sequential Pattern Mining and Support Vector Machine.

Using machine learning and data mining methods can extract the features of network data effectively, so compared with the traditional network analysis methods, they can obtain more satisfactory detection results. However, these methods require a large amount of data when training the model. The larger the amount of data, the better the classification effect. The network data currently used can be obtained in the following ways. 

\textbf{Packet-Level Data:} There are many protocols in the network, such as Transmission Control Protocol (TCP), User Datagram Protocol (UDP), Internet Control Message Protocol (ICMP), Internet Gateway Management Protocol (IGMP), etc. Users running these protocols generate the packet network traffic of the network. The network packets can be captured by a specific application programming interface (API) called pcap. Libpcap and WinPCap are the capture software libraries of many network tools, including protocol analyzers, packet sniffers, network monitors, network IDSs, and traffic generators. 

\textbf{NetFlow Data:} Ciscos NetFlow version 5 defines a network flow as a unidirectional sequence of packets with seven attributes: ingress interface, source IP address, destination IP address, IP protocol, source port, destination port, and IP type of service. Currently, there are 10 versions of NetFlow. Versions 1 to 8 are similar, but version 9 and version 10 have an important difference.

\textbf{Public Data Sets:} Some public data sets are commonly recognized and widely used in intrusion detection research. The Defense Advanced Research Projects Agency (DARPA) collected the datasets with the Massachusetts Institute of Technology Lincoln Laboratory (MIT/LL) in 1998\footnote{https://www.ll.mit.edu/r-d/datasets/1998-darpa-intrusion-detection-evaluation-dataset}. The later DARPA 1999 datasets\footnote{https://www.ll.mit.edu/r-d/datasets/1999-darpa-intrusion-detection-evaluation-dataset} and KDD 99 datasets~\cite{5356528} were generated on the DARPA 1998 datasets. These datasets lay the foundation for the application of machine learning and data mining methods in Intrusion Detection.

\subsection{Software Defined Networks (SDN)} 

The advantages of breaching the physical boundary, being programmable, and easily deploying make software-defined networks (SDN) become another promising solution for ICS network security. There are two types of flow control units in SDN, which are OpenFlow and optimalFlow respectively, which are named software controllers.

 Different ways to get ensure the security of the networks, some works think the security matters from the network structure of SDN. Manuel et al.~\cite{cheminod2017leveraging} designed an active SDN switch architecture that is acting as a virtual firewall. It accommodates different protocols and makes the management of the ICS network efficient. Bela et al.~\cite{genge2016hierarchical} redesigned the ILP problem over the SDN and built a hierarchical control plane over the ICS, and make the ICS networks more secure and flexible. Graur~\cite{graur2017dynamic} adopted the SDN developing a controller which can be applied in reconfiguring the ICS network. Some works adopted the IDS or IPS system besides the SDN to ensure security. Dong et al.~\cite{dong2015software} applied the IDS system along with the IDS to defend the attacks from the smart grid.

\section{Conclusions} \label{sec:future}
In this paper, we gave a comprehensive review of the recent research on ICS network security. We first listed the network protocols that are mostly used in the network and analyzed the probable vulnerabilities and their defense methods. Then, we try to give a brief review of the defense in depth in ICS in terms of data encryption, access control policy, and intrusion detection system. Finally, we ended with the software-defined network which is one of the promising research directions.



\bibliographystyle{IEEEtran}
\bibliography{ref} 
\end{document}